\documentclass[doublecol]{epl2}
\usepackage{flushend}

\title{Pd site doping effect on superconductivity in Nb$_{2}$Pd$_{0.76}$S$_{5}$}

\author{C. Y. Shen\inst{1} \and B. Q. Si\inst{1} \and H. Bai\inst{1} \and X. J. Yang\inst{1} \and Q. Tao\inst{1} \and G. H. Cao\inst{1,2,3} \and Z. A. Xu\inst{1,2,3}}

\institute{
  \inst{1} Department of Physics and State Key Laboratory of Silicon Materials, Zhejiang University, Hangzhou 310027, China

  \inst{2} Zhejiang California International NanoSystems Institute, Zhejiang University, Hangzhou 310058, China

  \inst{3} Collaborative Innovation Centre of Advanced Microstructures, Nanjing 210093, China
}


\abstract{ Pd site doping effect on superconductivity was
investigated in quasi-one-dimensional superconductor
Nb$_{2}$(Pd$_{1-x}$R$_{x}$)$_{0.76}$S$_{5}$ (R=Ir, Ag) by
measuring resistivity, magnetic susceptibility and Hall effect. It
was found that superconducting transition temperature ($T_c$) is
firstly slightly enhanced by partial substitution of Pd with Ir
and then it is suppressed gradually as Ir content increases
further. Meanwhile Ag substitution quickly suppresses the system
to a non-superconducting ground state. Hall effect measurements
indicate the variations of charge carrier density caused by Ir or
Ag doping. The established phase diagram implies that the charge
carrier density (or the band filling) could be one of the crucial
controlling factors to determine $T_c$ in this system.}

\begin{document}

\maketitle

\section{Introduction}

Recently superconductivity with $T_c$ of about 7 K has been
discovered in a transition-metal chalcogenide Nb$_{2}$PdS$_{5}$,
which displays extremely large upper critical field ($H_{c2}$),
violating the Pauli paramagnetic limit\cite{Puali} by a factor of
3\cite{Nb215,Nb215fiber}. In contrast to the iron pnictides, this
compound crystallizes in a lower symmetry space-group C2/m and was
argued to be a multi-band superconductor\cite{Nb215}. Further
investigation on the substitution effect shows that
superconductivity is still alive at $T_{c}$$\sim$6 K by replacing Nb
with Ta\cite{Ta215}, which is proposed to be close to the Anderson
localization state. On the other hand, $T_c$ is systematically
suppressed by partially substituting S by Se and eventually
superconductivity disappears with a semiconducting ground state at
50$\%$ Se substitution in Nb$_2$PdS$_{5-x}$Se$_x$ \cite{sedoped},
but $T_c$ is still as high as 2.5 K in the Ta-based system
Ta$_2$PdSe$_5$\cite{Tapdse215}. The large upper critical field
($H_{c2}$) is confirmed in all these
superconductors\cite{Nb215,Ta215,Nbpdse2x5,Tapdse215}, which
implies that it is a universal feature and these
quasi-one-dimensional (Q1D) compounds may belong to a new family
of unconventional superconductors.

The spin-orbit coupling (SOC) has been speculated to be a
potential ingredient for the unconventional properties in these
Q1D superconductors. The ratio of $H_{c2}$ to $T_c$ was reported
to be significantly enhanced by doping Pt into Nb$_2$PdS$_5$
system due to the strong spin-orbit coupling\cite{nb215niptdope}.
This is also consistent with the argument that the high $H_{c2}$
of Ta$_2$PdS$_5$ and Nb$_2$PdSe$_5$ derives from the large SOC of
heavy Pd element\cite{Nbpdse2x5,Ta215}. In addition, according to
the theoretical study of the electronic structure of these
compounds, 4$\emph{d}$ electrons of Pd contribute to the Fermi surface
mostly\cite{Nb215,Nbpdse2x5} and thus it is believed that the
heavy element Pd with 4$\emph{d}$ electrons should play a key role in the
mechanism of superconductivity\cite{Nb215,multiband}, but there are
very few studies on this issue.

In this Letter, we focus on the doping effect on Pd site by
heterovalent transition metals such as Ir and Ag. It turns out
that superconductivity can be slightly enhanced by partial
substitution of Pd by Ir, but is suppressed rapidly with Ag doping. A
phase diagram of Nb$_{2}$(Pd$_{1-x}$R$_{x}$)$_{0.76}$S$_{5}$
(R=Ir, Ag) is established, which indicates that superconductivity
is systematically affected by the 4$\emph{d}$ or 5$\emph{d}$ electron numbers on the
Pd site, and the charge carrier density could be one of the
crucial factors to control superconductivity in
Nb$_{2}$Pd$_{0.76}$S$_{5}$.

\section{Experiment}

\begin{figure}[htb]
\includegraphics[width=8cm]{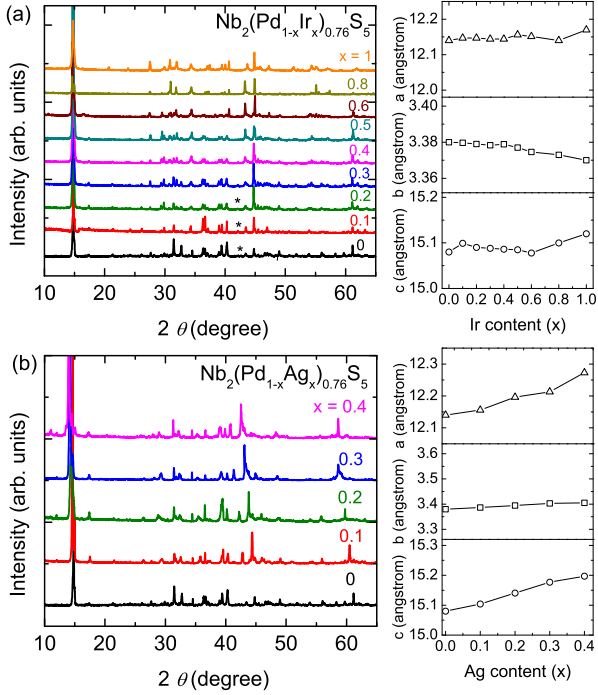}
\caption{(a) Powder X-ray diffraction patterns at room temperature
for Nb$_{2}$(Pd$_{1-x}$Ir$_{x}$)$_{0.76}$S$_{5}$ with $x$ from 0
to 1. (b) XRD patterns for
Nb$_{2}$(Pd$_{1-x}$Ag$_{x}$)$_{0.76}$S$_{5}$ with $x$ = 0, 0.1,
0.2, 0.3 and 0.4 respectively. The right panel plots Ir(Ag) doping
dependence of lattice parameters, where $a$, $b$ and $c$ are
denoted by triangle, square and circle symbols respectively. Minor
peaks from impurity are marked by asterisks.}
\end{figure}

Poly-crystalline samples of Nb$_{2}$Pd$_{1-x}$R$_{x}$S$_{5}$
(R=Ir, Ag) were synthesized by a solid-state reaction method.
Powders of Nb(99.99$\%$), Pd(99.99$\%$), Ir or Ag (99.99$\%$), and
S(99.9$\%$) were mixed and pelletized in a stoichiometric ratio of
2:1-$x$:$x$:6. The pellets were then sealed in an evacuated quartz
tube followed by a sintering procedure at 1123 K for 24 h. All
procedures were performed in a glove box filled with high-purity
argon except for the sintering process. The obtained samples usually
shows Pd site deficiency, which were found to be around 0.24
according to the measurements of energy-dispersive x-ray
spectroscopy (EDX). Namely the samples can be described as
Nb$_{2}$(Pd$_{1-x}$R$_{x}$)$_{0.76}$S$_{5}$. Both area (of about 4$\mu m^2$) and spot scans were used in the EDX measurements and the estimated errors are about 3\%-5\%.

Room temperature powder X-ray diffraction (XRD) was performed
using a PANalytical x-ray diffractometer (Model EMPYREAN) with a
monochromatic Cu-K$_{\alpha}$ radiation. The electrical
resistivity was measured using a standard four-terminal method.
The Hall effect was performed on a Quantum Design physical
property measurement system (PPMS-9). The DC magnetization
measurements were performed on a Quantum Design magnetic property
measurement system (MPMS-5). Both the zero field-cooling (ZFC) and
field-cooling (FC) measurements were applied.

\section{RESULTS AND DISCUSSION}

\begin{figure}[htb]
  \centering
  \includegraphics[width=8cm]{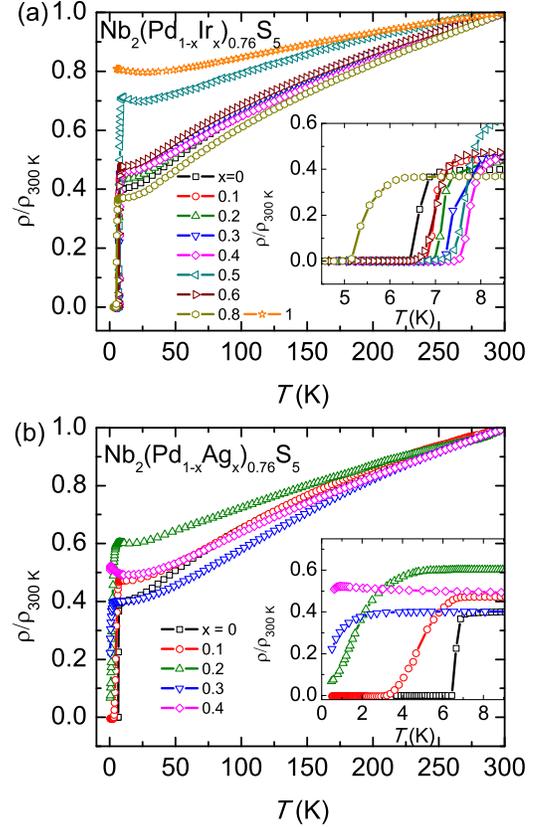}%
  \caption{(a) Temperature dependence of normalized resistivity for  Nb$_{2}$(Pd$_{1-x}$Ir$_{x}$)$_{0.76}$S$_{5}$ with $x$ from 0 to 1. (b) Temperature dependence of normalized resistivity for  Nb$_{2}$(Pd$_{1-x}$Ag$_{x}$)$_{0.76}$S$_{5}$ with $x$ = 0, 0.1, 0.2, 0.3 and 0.4 respectively. The insets zoom in the low temperature regimes below 8 K.}
\end{figure}
\begin{figure}[htb]
  \centering
  \includegraphics[width=8cm]{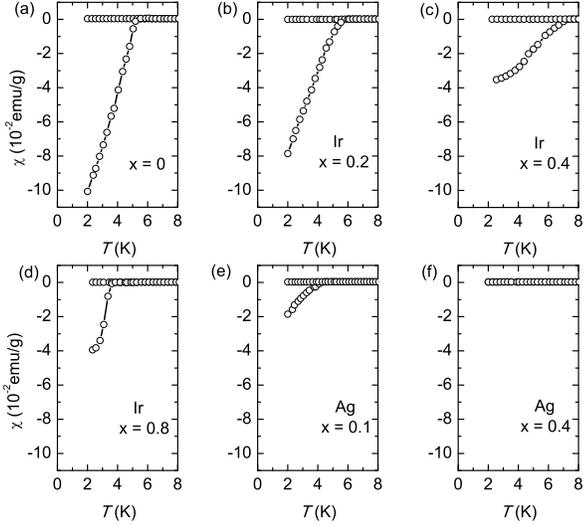}
  \caption{Temperature dependence of the dc magnetic susceptibility under ZFC and FC protocols for Nb$_{2}$(Pd$_{1-x}$Ir$_{x}$)$_{0.76}$S$_{5}$ and  Nb$_{2}$(Pd$_{1-x}$Ag$_{x}$)$_{0.76}$S$_{5}$ samples under a magnetic field of 10 Oe below 8 K.}
\end{figure}

Figure 1 (a) and (b) show the XRD patterns for the samples of
Nb$_{2}$(Pd$_{1-x}$Ir$_{x}$)$_{0.76}$S$_{5}$ and
Nb$_{2}$(Pd$_{1-x}$Ag$_{x}$)$_{0.76}$S$_{5}$ respectively. The
main XRD peaks can be well indexed based on the C2/m space group
except a few minor peaks assigned as impurity phase marked by the
asterisks, which is unknown yet. The right panels plot the lattice
parameters as a function of nominal Ir or Ag content. It is found
that for Ir doping, the lattice parameters seem less dependent on
the Ir content due to the comparable ion radius of Ir and Pd. In
contrast, the $a$ and $c$ axis for the Ag-doped samples increase
monotonically, consisting with the large radius of Ag ion, while
the $b$ axis does not change so much.

The temperature dependence of resistivity for
Nb$_{2}$(Pd$_{1-x}$Ir$_{x}$)$_{0.76}$S$_{5}$ and
Nb$_{2}$(Pd$_{1-x}$Ag$_{x}$)$_{0.76}$S$_{5}$ is presented in fig.
2. Each curve has been normalized by its room temperature value
for easy presentation. The absolute values of resistivity at 300 K range from 1.21 $m\Omega\cdot cm$ to 3.09 $m\Omega\cdot cm$,comparable with that found in Nb$_2$Pd$_x$Se$_5$\cite{Nbpdse2x5}. For the undoped compound
Nb$_2$Pd$_{0.76}$S$_5$, $\rho(T)$ remains metallic and becomes
superconducting below 6.8 K. Upon doping, a small upturn can be
seen in $\rho(T)$ at low temperatures, which is also found in Ta
or Se doped samples due to the proximity to an Anderson
localization associated with strong disorders\cite{Ta215,sedoped}. Such a small upturn in resistivity could also be caused by the grain boundary effect in polycrystalline samples, as reported previously in the high-$T_c$ cuprates\cite{boundary}.
Meanwhile, the critical temperature $T_{c}$ starts to increases
with Ir doping and reaches a maximum of 8 K at $x$ = 0.4, then
$T_c$ rapidly decreases and drops below 0.5 K at $x$ = 1. While
for Nb$_{2}$(Pd$_{1-x}$Ag$_{x}$)$_{0.76}$S$_{5}$, the
superconductivity is suppressed quickly with increasing Ag content
and finally disappears (for $T>0.5 K$) with 40$\%$ substitution of
Ag. The bulk nature of the superconductivity for the doping
samples was confirmed by the magnetization measurements, as given
in fig. 3, where large diamagnetic signals can be clearly seen for
the superconducting samples.

Figure 4 shows the Hall coefficient ($R_H$) as a function of
temperature for Nb$_{2}$(Pd$_{1-x}$Ir$_{x}$)$_{0.76}$S$_{5}$ and
Nb$_{2}$(Pd$_{1-x}$Ag$_{x}$)$_{0.76}$S$_{5}$ samples respectively. It may be noted that the Hall coefficient is usually not very sensitive to grain boundaries\cite{boundary,boundaryhall}. The $R_H$ of Nb$_{2}$(Pd$_{1-x}$Ir$_{x}$)$_{0.76}$S$_{5}$ is
positive in normal state, suggesting the dominant charge transport
by the hole conduction. This positive $R_H$ confirms the hole
doping by substituting Pd with the Ir element. On the other hand, the
negative Hall coefficient for
Nb$_{2}$(Pd$_{1-x}$Ag$_{x}$)$_{0.76}$S$_{5}$ implies the electron
dominant transport properties in the Ag doped samples, which is in
agreement with the additional electrons induced by replacing Pd
with Ag. The normal state $R_H$ ($T$) exhibits very weak
temperature dependence, which is usually observed in conventional
one-band metals. For a multi-band system, a strong temperature
dependence of $R_H$ is often expected. In addition, the linear magnetic field dependence of the Hall resistivity found in Nb$_2$Pd$_{1.2}$Se$_5$\cite{Nbpdse2x5} is also observed in our samples, suggesting that the Hall effect could be dominated by only one type of charge carriers.

\begin{figure}[htb]
  \centering
  \includegraphics[width=8cm]{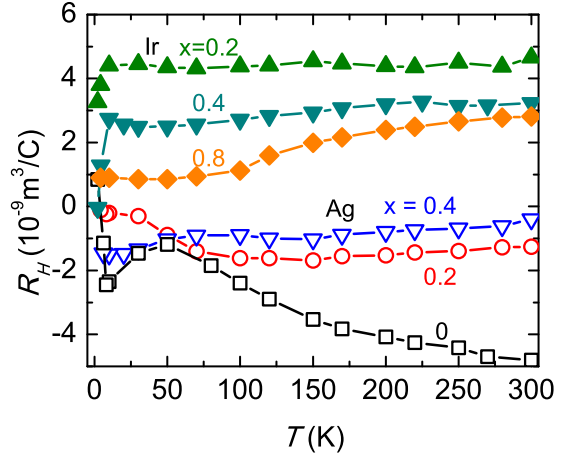}
  \caption{Temperature dependence of the Hall coefficient for Nb$_{2}$(Pd$_{1-x}$Ir$_{x}$)$_{0.76}$S$_{5}$ (full symbols) and  Nb$_{2}$(Pd$_{1-x}$Ag$_{x}$)$_{0.76}$S$_{5}$ samples (empty symbols) respectively.}
\end{figure}
\begin{figure}[htb]
  \centering
  \includegraphics[width=8cm]{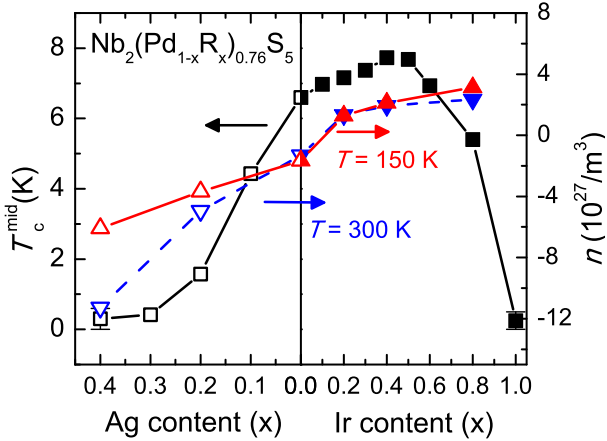}
  \caption{Phase diagram of $T_c^{mid}$ (square), $n_{T=150K}$ (upper triangle, solid line) and $n_{T=300K}$ (down triangle, dashed line) vs. doping for Nb$_{2}$(Pd$_{1-x}$Ir$_{x}$)$_{0.76}$S$_{5}$ (full symbols) and Nb$_{2}$(Pd$_{1-x}$Ag$_{x}$)$_{0.76}$S$_{5}$ (empty symbols). The negative (positive) value of $n_{T=150K}$ (or $n_{T=300K}$) indicates the electron (hole)-type charge carriers.}
\end{figure}

The middle point of superconducting transition temperature in
resistivity, $T_c^{mid}$, for
Nb$_{2}$(Pd$_{1-x}$R$_{x}$)$_{0.76}$S$_{5}$ (R=Ir, Ag) as well as
the estimated charge carrier density ($n$) extracted from $R_H$ at
$T$ = 150 K and 300 K is plotted against Ag (Ir) content ($x$) in fig. 5.
Since the Hall effect might be dominated by one band, we just simply assume a single band model to estimate the carrier density using $R_H=\frac{1}{ne}$. Although the carrier density extracted through a single band model is not very accurate considering the complex Fermi surfaces in this system, it could still give a qualitative estimation on the evolution of carrier numbers with the doping. Giving the heterovalent
doping of Ir and Ag, it is believed that such dopants will change
the carrier density in the system, which is indeed confirmed by
the Hall measurements. As shown in fig. 5, a monotonic increase of
the carrier density with increasing Ir (Ag) doping is clearly
observed, although the change rate of carrier density with doping
content ($x$) for the two dopants are slightly different. Overall,
superconductivity is firstly enhanced by the hole-type doping (Ir
doping) and $T_c^{mid}$ reaches the highest value ($\sim$ 8 K) around
$x$ (Ir) = 0.4. But upon further hole doping, $T_c^{mid}$ is
rapidly suppressed. On the contrast, with electron-type doping (Ag
doping), $T_c^{mid}$ decreases monotonically with $x$ (i.e.,
increasing electron carrier density), and finally
superconductivity disappears (for $T$ $>$ 0.5 K), suggesting a
significant negative correlation between electron-type carrier
density and superconductivity. Thus our data illustrate that the
suitable charge carrier density (or band filling) could be crucial
to the occurrence of superconductivity.

The observation above could be understood in the following
scenario. According to the band
calculations\cite{Nb215,Nbpdse2x5}, the parent compound
Nb$_{2}$PdS$_{5}$ has been predicted to be very close to an
itinerant magnetically ordered state with a complex Fermi surface
(FS) composed of Quasi-two-dimensional (Q2D) sheets of hole
character and strongly warped quasi-one-dimensional sheets of both
hole and electron. By adding a small amount of extra negative
(electron) charge into the system, the very flat band near Fermi
energy will become the Q1D Fermi surface sheets, giving rise to
strong nesting properties as well as large density states, so that
the long range magnetic order could be stabilized. It is well
known that, in many unconventional superconductors, the magnetic
order such as anti-ferromagnetism and spin density wave (SDW)
competes with
superconductivity\cite{cupratesmagnetic,ironbasedmagnetic}.
Therefore the negative correlation between electron-type charge
carrier density and superconductivity in Nb$_2$Pd$_{0.76}$S$_5$
could be well understood in this manner. Namely Ag doping induces
electron-type charge carriers to the system and then drives it
even close to the magnetic instability. Another possible scenario is that the strong nest properties of FS may allow the formation of charge density wave (CDW) in low-dimensional systems, which usually competes with superconductivity\cite{Nbpdse2x5}. In this case, Ag doping may drive the system close to the CDW order. Nevertheless, the
situation for the hole doping (Ir doping) is somewhat complicated.
$T_c$ initially increase with hole-type charge carrier density,
reaches a maximum of about 8 K, then decreases quickly with
further hole-type doping. Although we can naively assume that the
hole-type doping could drive the system away from the magnetic
instability and thus it has a positive effect on
superconductivity, the reason why $T_c$ drops with further Ir
doping remains an open issue. Note that the chemical pressure, for
example, the substitution of S by Se\cite{sedoped}, could reduce
$T_c$ significantly in this Q1D system. However, in the case of
Ir-for-Pd doping, the lattice parameters do not change so much,
thus the effect of chemical pressure might be ruled out. Another
possible explanation may be related to the variation in the
strength of the spin-orbit coupling due to Ir doping. Since Ir has
larger atomic number than Pd, its inherent SOC should be naturally
larger than that of Pd. Therefore, with increasing Ir doping, the
strength of SOC could be enhanced. However, recent experimental
studies on superconductivity of the LaAlO$_3$/SrTiO$_3$ interface
and the transport properties in a Pb thin film under an in-plane
magnetic field show that large SOC has a positive impact on
low-dimensional superconductivity\cite{socon2D,SOCinptfilms}.
Moreover, a theoretical study on Bi-rich compounds ABi$_3$ (A=Sr
and Ba)\cite{ABe3}, which are of a three-dimensional structure,
demonstrates that superconductivity could be significantly
enhanced due to the phonon softening and an increase in
electron-phonon coupling induced by SOC. Therefore, the effect of
SOC on $T_c$ in Nb$_2$PdS$_5$ system is an interesting issue. In
the case of Pt doping, in contrast to Ir doping, $T_c$ is
depressed, but its SOC should be enhanced\cite{nb215niptdope} as
in the Ir-doping case.

An interesting feature of the phase diagram shown in fig. 5 is
that the dependence of $T_c$ on the charge carrier density is also
dome-like, which mimics the phase diagram in high-$T_c$
superconductors such as the cuprates\cite{Revcuprate} and iron
pinctides\cite{RevIron} despite of the sign change in the dominant
charge carriers. Upon electron (hole) doping, superconductivity
emerges and $T_c$ increases, reaches a maximum at an ``optimal
doping level'' at $x$ (Ir) = 0.4, and then it is suppressed
quickly. Finally superconductivity disappears in the so-called
"overdoped region". Giving that the parents compounds of cuprates
and iron pnictides are antiferromagnetically (AFM)
ordered\cite{AFMcuprate,AFMiron}, the proximity to a long range
magnetic order in the Nb$_2$PdS$_5$ system predicated by the band
calculations, though not experimentally observed so far, implies
the similarity in phase diagrams for these unconventional
superconductors, which may imply that the mechanism of
superconductivity of the Nb$_2$PdS$_5$ system could have a close
relationship with these systems. On the other hand,
superconductivity is often found close to a quantum critical point
(QCP) in the materials where a long-range magnetic order is
gradually suppressed as a function of a control parameter such as
charge-carrier doping or pressure. It is an interesting issue if
there exists a parent compound with magnetic order in the
Nb$_2$PdS$_5$ system and a magnetic QCP accompanied by the emergence
of superconductivity. Nb$_2$Pd$_{0.76}$S$_5$ is found to be
Fermi-liquid like in the temperature range just above $T_c$,
however, other chalcogenides such as
Nb$_2$Pd$_x$Se$_5$\cite{Nbpdse2x5} and
Nb$_3$Pd$_x$Se$_7$\cite{multiband} display metallic state with
non-Fermi-liquid behavior at very low temperatures. To elucidate
such issues, more theoretical and experimental studies are
required.

\section{CONCLUSION}
In summary, we have investigated the superconducting properties in
the Nb$_{2}$(Pd$_{1-x}$R$_{x}$)$_{0.76}$S$_{5}$ (R=Ir, Ag)
polycrystalline samples. It is found that  superconductivity can
be enhanced by partial substitution of Ir but is quickly suppressed
to a nonsuperconducting ground state with 40$\%$ Ag doping. Ag
doping is an electron-type dopant and Ir doping could be regarded
as the hole-type dopant as suggested by the Hall effect
measurements. The overall phase diagram indicates a domelike
dependence of $T_c$ on the doping level, which mimics the general
phase diagrams of high-$T_c$ superconducting cuprates and some
other unconventional superconductors. Our work implies that there could exist a competing order, either magnetic order or charge order in this system and the charge carrier density (or band filling) is one of the crucial factors to
tune the superconducting order.

\acknowledgments
The work is supported by the National Basic Research Program of China
(Grant Nos. 2014CB921203 and 2012CB821404), the National Science
Foundation of China (Grant Nos. 11190023, U1332209 and 11174247),
and the Fundamental Research Funds for the Central Universities of
China.


\begin{thebibliography}{10}
\expandafter\ifx\csname url\endcsname\relax\def\url#1{\texttt{#1}}\fi

\bibitem{Puali}
\Name{Clogston A.} \REVIEW{Phys. Rev. Lett.}{9}{1962}{266}.

\bibitem{Nb215}
\Name{Zhang Q., Li G., Rhodes D., Kiswandhi A., Besara T., Zeng B., Sun J.,
  Siegrist T., Johannes M. \and Balicas L.} \REVIEW{Sci. Rep.}{3}{2013}{1446}.

\bibitem{Nb215fiber}
\Name{Yu H.-y., Zuo M., Zhang L., Tan S., Zhang C.-J. \and Zhang Y.-H.}
  \REVIEW{J. Am. Chem. Soc.}{135}{2013}{12987}.

\bibitem{Ta215}
\Name{Lu Y.-F., Takayama T., Bangura A.~F., Katsura Y., Hashizume D. \and
  Takagi H.} \REVIEW{J. Phys. Soc. Jpn.}{83}{2013}{023702}.

\bibitem{sedoped}
\Name{Niu C.-Q., Yang J.-H., Li Y.-K., Chen B., Zhou N., Chen J., Jiang L.-L.,
  Chen B., Yang X.-X., Cao C., Dai J.-H. \and Xu X.-F.} \REVIEW{Phys. Rev.
  B}{88}{2013}{104507}.

\bibitem{Tapdse215}
\Name{Zhang J., Dong J.-K., Xu Y., Pan J., He L.-P., Zhang L.-J. \and Li S.-Y.}
  \REVIEW{Supercond. Sci. Technol.}{28}{2015}{115015}.

\bibitem{Nbpdse2x5}
\Name{Khim S., Lee B., Choi K.-Y., Jeon B.-G., Jang D.~H., Patil D., Patil S.,
  Kim R., Choi E.~S., Lee S. \etal} \REVIEW{New J. Phys.}{15}{2013}{123031}.

\bibitem{nb215niptdope}
\Name{Zhou N., Xu X.-F., Wang J.-R., Yang J.-H., Li Y.-K., Guo Y., Yang W.-Z.,
  Niu C.-Q., Chen B., Cao C. \and Dai J.-H.} \REVIEW{Phys. Rev.
  B}{90}{2014}{094520}.

\bibitem{multiband}
\Name{Zhang Q.-R., Rhodes D., Zeng B., Besara T., Siegrist T., Johannes M. D.
  \and Balicas L.} \REVIEW{Phys. Rev. B}{88}{2013}{024508}.

\bibitem{boundary}
\Name{Carrington A. \and Cooper J. R.} \REVIEW{Physica C}{219}{1994}{119}.

\bibitem{boundaryhall}
\Name{Carrington A., Mackenzie A. P., Lin C. T. \and Cooper J. R.} \REVIEW{Phys. Rev. Lett.}{69}{1992}{2855}.

\bibitem{cupratesmagnetic}
\Name{S{\'e}n{\'e}chal D., Lavertu P.-L., Marois M.-A. \and Tremblay A.-M.}
  \REVIEW{Phys. Rev. Lett.}{94}{2005}{156404}.

\bibitem{ironbasedmagnetic}
\Name{de~La~Cruz C., Huang Q., Lynn J., Li J., Ratcliff~II W., Zarestky J.~L.,
  Mook H., Chen G., Luo J., Wang N. \etal} \REVIEW{Nature}{453}{2008}{899}.
\bibitem{socon2D}
\Name{Caviglia A.~D., Gabay M., Gariglio S., Reyren N., Cancellieri C. \and
  Triscone J.-M.} \REVIEW{Phys. Rev. Lett.}{104}{2010}{126803}.

\bibitem{SOCinptfilms}
\Name{Gardner H.~J., Kumar A., Yu L., Xiong P., Warusawithana M.~P., Wang L.,
  Vafek O. \and Schlom D.~G.} \REVIEW{Nat. Phys.}{7}{2011}{895}.

\bibitem{ABe3}
\Name{Shao D., Luo X., Lu W., Hu L., Zhu X., Song W., Zhu X. \and Sun Y.}
  \REVIEW{arXiv preprint arXiv:1510.05719}{}{2015}{}.

\bibitem{Revcuprate}
\Name{Lee P.~A., Nagaosa N. \and Wen X.-G.}
\REVIEW{Rev. Mod.
  Phys.}{78}{2006}{17}.

\bibitem{RevIron}
\Name{Stewart G.~R.} \REVIEW{Rev. Mod.
Phys.}{83}{2011}{1589}.

\bibitem{AFMcuprate}
\Name{Vaknin D., Sinha S.~K., Moncton D.~E., Johnston D.~C., Newsam J.~M.,
  Safinya C.~R. \and King H.~E.} \REVIEW{Phys. Rev. Lett.}{58}{1987}{2802}.

\bibitem{AFMiron}
\Name{Huang Q., Qiu Y., Bao W., Green M.~A., Lynn J.~W., Gasparovic Y.~C., Wu
  T., Wu G. \and Chen X.~H.} \REVIEW{Phys. Rev. Lett.}{101}{2008}{257003}.

\end{thebibliography}
\end{document}